
  \input epsf
\input harvmac
\noblackbox
\def\Title#1#2{\rightline{#1}\ifx\answ\bigans\nopagenumbers\pageno0\vskip1in
\else\pageno1\vskip.8in\fi \centerline{\titlefont #2}\vskip .5in}

%
%
\ifx\epsfbox\UnDeFiNeD\message{(NO epsf.tex, FIGURES WILL BE IGNORED)}
\def\figin#1{\vskip2in}
\else\message{(FIGURES WILL BE INCLUDED)}\def\figin#1{#1}
\fi
\def\Fig#1{Fig.~\the\figno\xdef#1{Fig.~\the\figno}\global\advance\figno
 by1}
%
%
%
%
\def\ifig#1#2#3#4{
\goodbreak\midinsert
\figin{\centerline{\epsfysize=#4truein\epsfbox{#3}}}
\narrower\narrower\noindent{\footnotefont
{\bf #1:}  #2\par}
\endinsert
}
%
%
\def\jou#1&#2(#3){\unskip, \sl#1\bf#2\rm(19#3)}
\def\ajou#1&#2(#3){\ \sl#1\bf#2\rm(19#3)}
\def\gtwid{\mathrel{\raise.3ex\hbox{$>$\kern-.75em\lower1ex\hbox{$\sim$}}}}
\def\ltwid{\mathrel{\raise.3ex\hbox{$<$\kern-.75em\lower1ex\hbox{$\sim$}}}}

\font\ticp=cmcsc10
%
%
\lref\BaOl{T. Banks and M. O'Loughlin, ``Classical and quantum production
of cornucopions at energies below $10^{18}$ GeV,'' hep-th/9206055
\jou Phys.Rev. &D47 (93) 540-553.}
\lref\BOS{T. Banks, M. O'Loughlin, and A. Strominger, ``Black hole remnants
and the information puzzle,'' hep-th/9211030 \jou Phys. Rev.& D47 (93) 4476.}
\lref\GiTa{S.B. Giddings and T. Tada, in progress.}
\lref\SVV{K. Schoutens, H. Verlinde, and  E. Verlinde, ``Quantum
black hole evaporation,''  hep-th/9304128\jou Phys. Rev. &D48 (93) 2670.}
\lref\tHoo{G. 't Hooft, ``Fundamental aspects of quantum theory
related to
the problem of quantizing black holes,''  in {\sl Quantum Coherence},
ed.
J.S. Anandan (World Sci., 1990), and references therein\semi
Talk at 1993 ITP Conference, Quantum
Aspects of Black Holes.}
\lref\GiHa{G.W. Gibbons and S.W. Hawking, ``Action integrals and partition
functions in quantum gravity,''\ajou Phys. Rev. &D15 (77) 2752.}
\lref\SUSSTFR{L. Susskind, ``Trouble for remnants,'' Stanford preprint
SU-ITP-95-1, hep-th/9501106.}
\lref\Schw{J. Schwinger, ``On gauge invariance and vacuum
polarization,''\ajou Phys. Rev. &82 (51) 664.}
\lref\Sussetal{L. Susskind, L. Thorlacius, and
J. Uglum, ``The stretched horizon
and black hole complementarity,'' hep-th/9306069
\jou Phys. Rev. &D48 (93) 3743\semi
L. Susskind, ``String theory and the principles of black
hole
complementarity,'' hep-th/9307168\jou Phys. Rev. Lett. &71 (93) 2367\semi
L. Susskind and L. Thorlacius, ``Gedanken experiments involving black
holes,''  hep-th/9308100\jou Phys. Rev. &D49 (94) 966\semi
L. Susskind, ``Strings, black holes and Lorentz contraction,''
hep-th/930813\jou Phys. Rev. &D49 (94) 6606.}
\lref\GiNe{S.B. Giddings and W.M. Nelson, ``Quantum emission from
two-dimensional black holes,'' hep-th/9204072 \jou Phys. Rev. &D46 (92)
2486.}
\lref\Hawk{S.W. Hawking, ``Particle creation by black
holes,"\ajou Comm. Math. Phys. &43 (75) 199.}
\lref\BPS{T. Banks, M.E. Peskin, and L. Susskind, ``Difficulties for
the evolution of pure states into mixed states,''\ajou Nucl. Phys. &B244
(84) 125.}
\lref\Sred{M. Srednicki, ``Is purity eternal?,'' hep-th/920605 \jou
Nucl. Phys. &B410 (93) 143.}
\lref\SBGTr{S.B. Giddings, ``Quantum mechanics of black holes,''
hep-th/9412138, to appear in the proceedings of the 1994 Trieste Summer
School in High Energy Physics and Cosmology.}
\lref\Cole{S. Coleman, ``Black Holes as Red Herrings:
Topological Fluctuations and the Loss of Quantum Coherence,''\ajou Nucl.
Phys. & B307 (88) 864.}
\lref\GiSt{S.B. Giddings and A. Strominger, ``Loss
of Incoherence and Determination of Coupling Constants
in Quantum Gravity,"\ajou Nucl. Phys. &B307 (88) 854.}
\lref\CaWi{R.D. Carlitz and R.S. Willey, ``Reflections on moving
mirrors,''\ajou Phys. Rev. &D36 (87) 2327; ``Lifetime of a black
hole,''\ajou Phys. Rev. &D36 (87) 2336.}
\lref\Pres{J. Preskill, ``Do black holes destroy information?''
hep-th/9209058, in the
proceedings of the International Symposium on Black holes, Membranes,
Wormholes and Superstrings,
Woodlands, TX, 16-18 Jan 1992.}
\lref\WABHIP{S.B. Giddings, ``Why aren't black holes infinitely
produced?,'' hep-th/9412159\jou Phys. Rev. &D51 (95) 6860.}
\lref\StTr{A. Strominger and S. Trivedi, ``Information consumption by
Reissner-Nordstrom black holes,'' hep-th/9302080
\jou Phys. Rev. & D48 (93) 5778.}
\lref\Ernst{F. J. Ernst, ``Removal of the nodal singularity of the
C-metric,'' \ajou J. Math. Phys. &17 (76) 515.}
\lref\Gibb{G.W. Gibbons, ``Quantized flux tubes in Einstein-Maxwell
theory and  noncompact internal
spaces,'' in {\sl Fields and geometry}, proceedings of
22nd Karpacz Winter School of Theoretical Physics: Fields and
Geometry, Karpacz, Poland, Feb 17 - Mar 1, 1986, ed. A. Jadczyk
(World Scientific, 1986).}
\lref\GaSt{D. Garfinkle and A. Strominger, ``Semiclassical Wheeler
wormhole production,''\ajou Phys. Lett. &B256 (91) 146.}
\lref\GGS{D. Garfinkle, S.B. Giddings, and A. Strominger, ``Entropy in
Black Hole Pair Production,'' gr-qc/9306023 \jou Phys. Rev. &D49 (94) 958.}
\lref\PSTU{D.A. Lowe et. al., ``Black hole complementarity versus
locality,'' ITP preprint NSF-ITP-95-47, hep-th/9506138.}
\lref\Pol{J. Polchinski, ``String theory and black hole complementarity,''
ITP preprint NSF-ITP-95-63, hep-th/9507094, to appear in the Proceedings of
Strings '95.}
\Title{\vbox{\baselineskip12pt\hbox{UCSBTH-95-24}\hbox{hep-th/9508151}
}}
{\vbox{\centerline {The Black Hole Information Paradox}
}}
\centerline{{\ticp Steven B. Giddings}\footnote{$^\dagger$}
{Email address:
giddings@denali.physics.ucsb.edu}
}
\vskip.1in
\centerline{\sl Department of Physics}
\centerline{\sl University of California}
\centerline{\sl Santa Barbara, CA 93106-9530}

\bigskip
\centerline{\bf Abstract}

A concise survey of the black hole information paradox and its current
status is given. A
summary is also given of recent arguments against remnants. The
assumptions underlying remnants, namely unitarity
and causality, would imply that Reissner Nordstrom black holes have
infinite internal states.  These can be argued to lead to an unacceptable
infinite production rate of such black holes in background fields.

(To appear in the proceedings of the PASCOS
symposium/Johns Hopkins Workshop, Baltimore, MD, March 22-25, 1995).

\Date{}

Black holes exist--we have seen new experimental evidence for this
claim in descriptions of Hubble telescope data at this conference.
In my talk I'd like to explain to you how this simple statement gets us
into a deep paradox in our understanding of nature. This paradox is
pushing us to consider abandoning one of our cherished principles of
energy conservation, locality and causality, or stability. It suggests
that a revision of the fundamental underpinnings of physics may be
necessary.

To see what the problem is, consider the formation of a large black hole
from a collapsing star of mass ${M_0}$,
as shown in fig.~1. Once it has
formed, Hawking \refs{\Hawk} has shown that, according to the basic
principles of quantum field theory, it will radiate its mass away.  This
result can be understood by recalling that in quantum field theory the
vacuum is populated by virtual pairs of particles and antiparticles.  The
gravitational field of the black hole can promote one member of
a pair into a real outgoing particle, while the other partner falls into
the black hole with a net negative energy as measured at infinity.

\ifig{\Fig\infall}{A spacetime diagram of collapse of a star to form a black
hole.  Hawking radiation arises from escape of one of a pair of virtual
particles, and leads to evaporation of the black hole.  Also shown are
examples of light cones.}{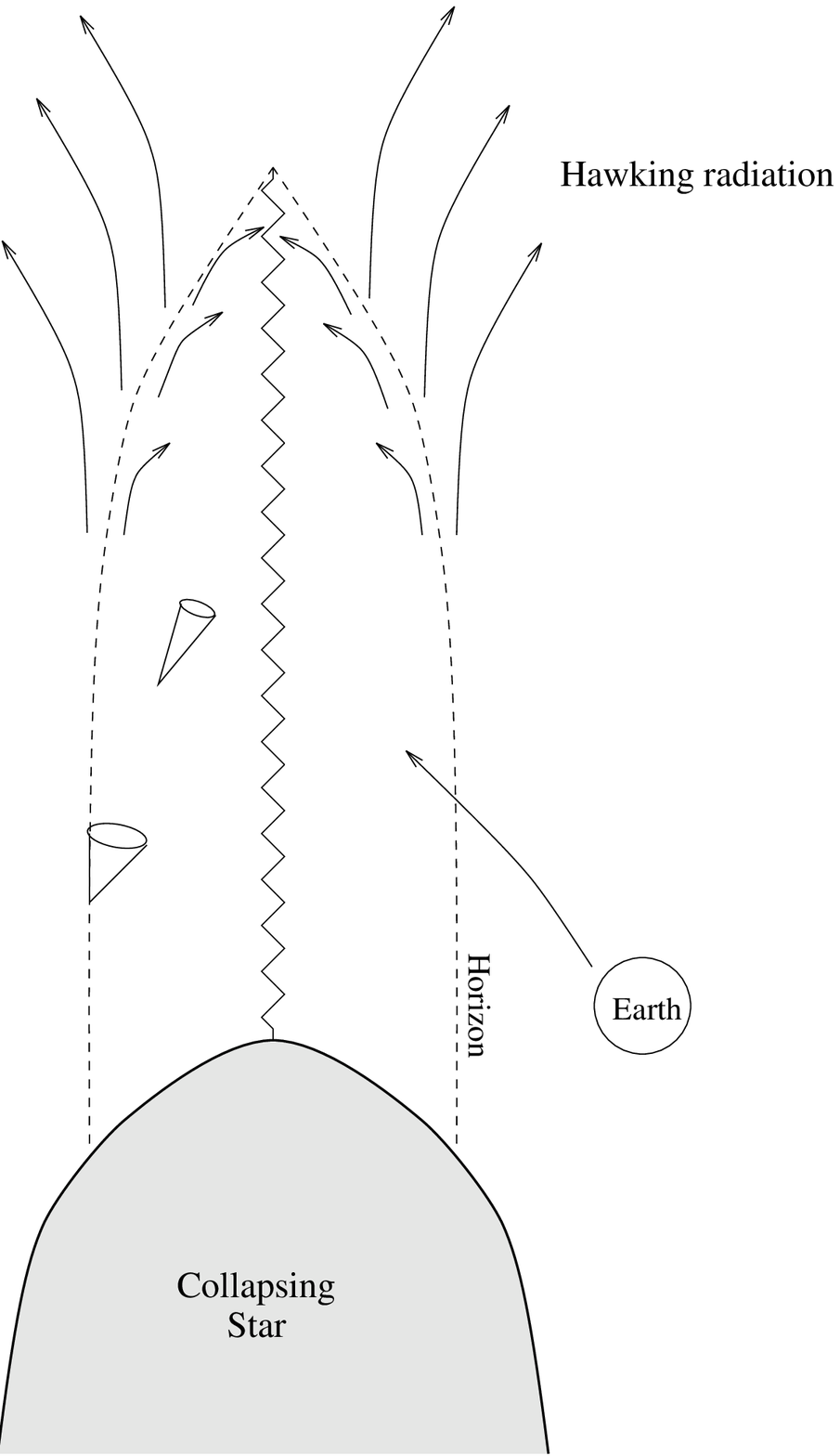}{4.0}

Now lets ask ourselves a question: suppose that the Earth were to fall
into this black hole, destroying our civilization. However, a
sufficiently advanced civilization might do very careful measurements on
the outgoing radiation from the black hole.  Could they be
reconstruct from these the history of our planet up to its
annihilation--would we at least leave some legacy?
%

To illustrate further, suppose the Earth were
blown to bits by a powerful
bomb. In that case, we know that, at least in principle, an advanced
civilization could decipher our history right up to the explosion--this
would of course require a very careful measurement of the outgoing
quantum state of the fragments, and then a backwards evolution of it to
the initial state before the explosion.

But according to Hawking, black holes are different. Here it would be
impossible even in principle to reconstruct the initial state. To see
this, consider the idealization where the initial state of the black
hole (including the Earth) was a quantum-mechanical pure state: in
density-matrix language,
\eqn\one{
{\rho} = |{\psi}{\rangle}{\langle}{\psi}|.}
This has zero entropy ${S}$, where
\eqn\two{
{S} = {-} {Tr} {\rho}{\ell}{n}{\rho}.}
Hawking showed that the outgoing radiation is approximately thermal, and
carries a large entropy
\eqn\three{
{S} {\sim} {M_0^2}/{M_{{p}{\ell}}^2},}
where ${M_{p\ell}}$ is the Planck mass. Correspondingly, once the black
hole has evaporated away completely the final state is a mixed state of
the form
\eqn\four{
{\rho} = {\sum_n}{p_n}|{n}{\rangle}{\langle}{n}|.}
(One can in fact derive such an expression directly\Hawk.)
Such evolution of a pure state to a mixed state implies a fundamental
loss of information
\eqn\infoloss{
{\Delta}{I} = {-}{\Delta}{S} {\sim} {-}{M^2_0}/{M_{p\ell}^2}.}
This loss of information ensures that not only we but also our history
would be obliterated.

As Hawking also pointed out, this behavior violates quantum mechanics,
which keeps pure states pure. Our state of
uneasiness about this is significantly
heightened when we realize that such information loss also apparently
implies energy non-conservation \refs{\BPS\Sred-\SBGTr}.

\ifig{\Fig\Vblack}{Shown is the contribution of a virtual black hole to two
body scattering; information and energy loss are expected to contaminate
arbitrary processes through such diagrams.}{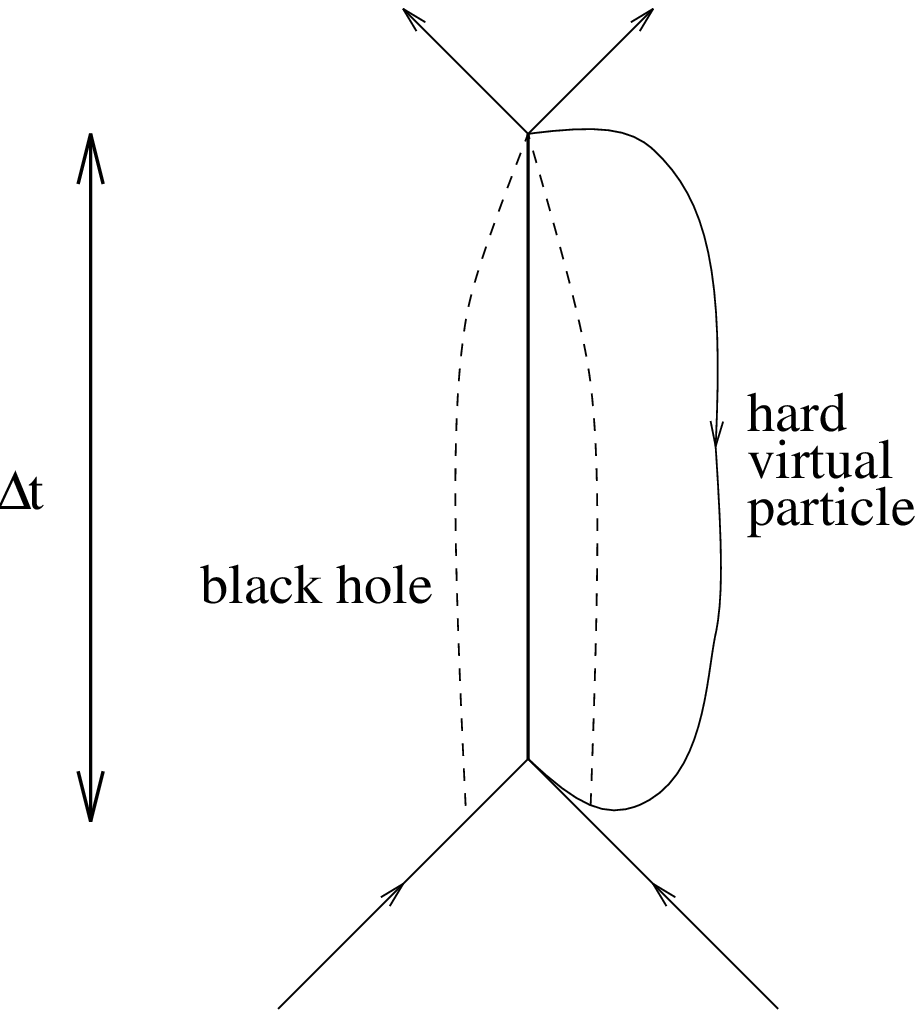}{2.00}

The basic point here is that information transmittal requires energy, so
its loss should imply loss of energy. Indeed, suppose that the black
hole forms and evaporates in a time ${\Delta}{t}$; then one expects an
uncertainty-principle like relation
\eqn\eninfo{
{\Delta}{E}{\sim}{1}/{\Delta}{t}}
governing the minimum amount of energy lost.\foot{There are examples of
information loss without energy loss, as illustrated by spacetime
wormholes \refs{\Cole,\GiSt}. However, such loss is not {\it repeatable}
\refs{\SBGTr}--it will not persist as needed to describe a
sequence of black hole formations
and evaporations.} Basic quantum principles imply that such
formation/evaporation should be taking place all the time in virtual
processes, as illustrated in fig.~2. The amplitude for these processes
should approach unity as the size of the loop approaches the Planck
scale--there is no small dimensionless number to suppress it. According
to \eninfo, we would therefore expect Planck size energy violations
with planckian characteristic time scale. This would give the world the
appearance of a thermal bath at the Planck temperature, in clear
contradiction with experiment. And that suggests we explore
alternatives to Hawking's picture.

\ifig{\Fig\Pencoll}{By a conformal transformation, fig.~1 can be redrawn in
the form of a Penrose diagram, in which light travels at 45$^\circ$
angles.}{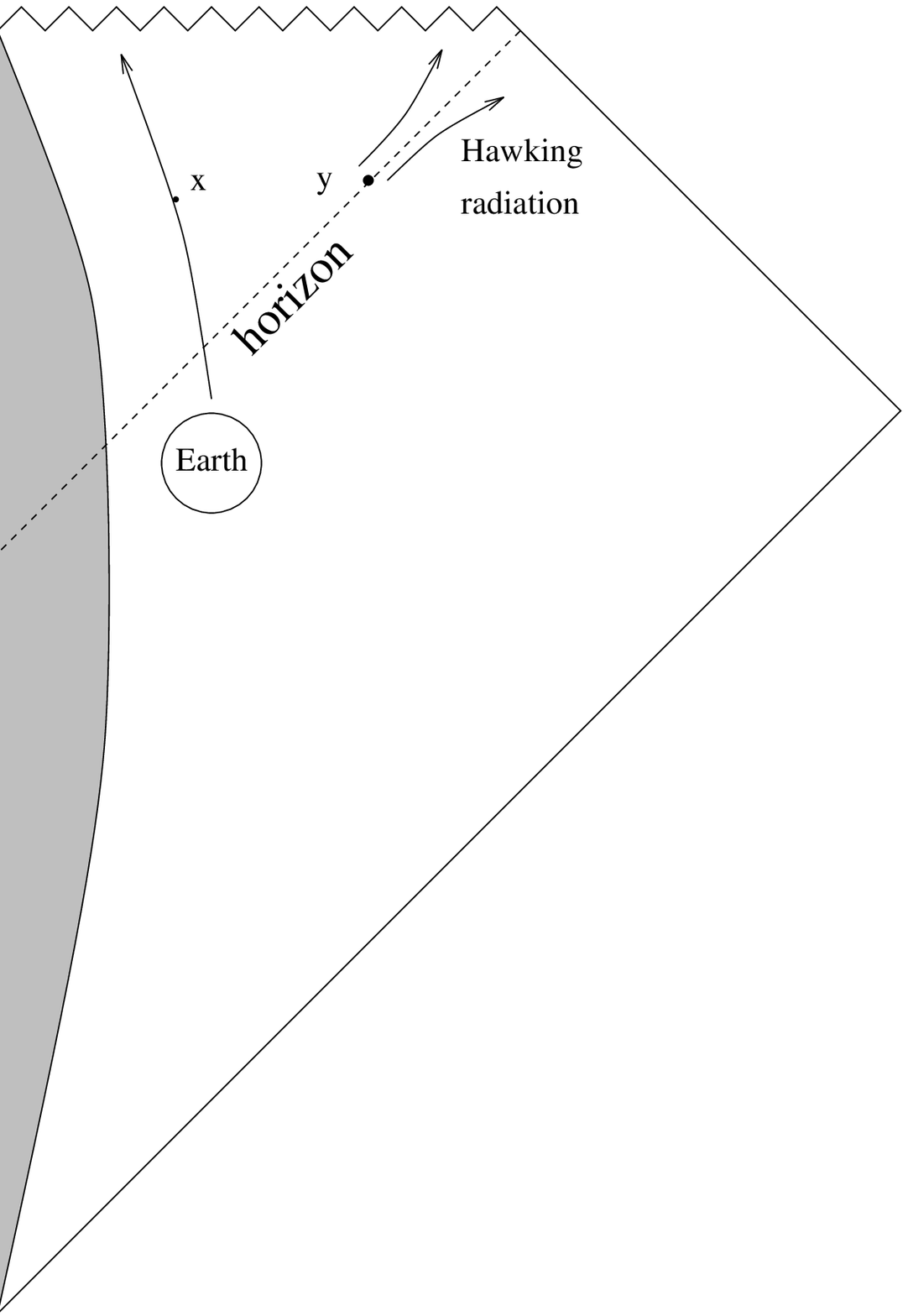}{4.00}

Can the information come out in the Hawking radiation? To investigate
this it is useful to redraw fig.~1 as a Penrose diagram, fig.~3. In such
a diagram light rays are at 45$^\circ$ angles, which makes causality
easy to study. From this diagram the difficulty is clear. The Hawking
radiation is produced near the event horizon at point y. For an event at
x in the history of the Earth to influence this, a signal would have to
be transmitted beyond the speed of light, that is acausally or
nonlocally. In field theory parlance, observables at x and y would have
to fail to commute,
\eqn\five{
[{\cal O}_1({x}), {\cal O}_2({y})]\,\,{\not=}\,\,{0},}
for spacelike separated x and y. Once we've allowed such violations of
locality/causality, we have to explain why they don't imply
paradoxes such as the possibility of
killing one's grandmother or winning money at the
racetrack.

This prompts consideration of other scenarios. Another apparent
possibility is that all of the information is released in the final
burst, as the black hole evaporates past the Planck mass. By this time
locality/causality inside the black hole has broken down. However, here
we confront another problem. The energy available is of order the Planck
mass, ${E} {\sim} {M_{p\ell}}$. The information that is to be radiated
is large, as we saw in eq.~\infoloss. Again, in accordance with
uncertainty principle arguments, the only way to radiate a large amount
of information with a small amount of available energy is to do it very
slowly, for example by emitting extremely soft photons. An estimate
\refs{\CaWi,\Pres} of the time required is
\eqn\six{
{t} {\sim} \biggl({{M_0}\over{M_{p\ell}}}\biggr)^4 {t_{p\ell}},}
which exceeds the age of the universe for black holes with initial
masses comparable to that of an average building. Therefore this
scenario implies long-lived remnants.

Long-lived black hole remnants thus remain as our final possibility.
These should have masses of order ${M_{p\ell}}$. There should
also be an infinite number of species of such objects, since they must
be able to encode in their internal states the information from an
arbitrarily large black hole. This leads to the final objection: in any
process with total available energy greater than ${M_{p\ell}}$ (e.g.
in nuclear reactors), there is a tiny but non-zero amplitude to
pair produce a given species of remnant. When multiplied by the infinite
number of species, this gives an infinite rate: the Universe is unstable
to instantaneous decay into remnants, again in clear contradiction with
experiment.

We have now painted ourselves into the corner that is the black hole
information paradox. Information loss implies energy non-conservation,
information in Hawking radiation requires non-locality and/or
acausality, and remnants lead to catastrophic instabilities. Beginning
with the observation that black holes exist and applying basic physical
principles has gotten us into serious trouble.

For a time it seemed that the most likely out was remnants--one might
imagine that planckian physics would allow us to find a loophole by
which they weren't infinitely produced, and proposals along these lines
have been made by \refs{\BaOl,\BOS}. This now appears to me quite
unlikely. Before concluding this talk with my views, I'll outline more
recent
reasoning that makes remnants look very unlikely to me. More details are
in \refs{\WABHIP}, and related objections are given in \refs{\SUSSTFR}.

The objection arises from considering black hole pair production, a
phenomenon of great interest in its own right. To see its relevance,
consider two properties of Reissner-Nordstrom black holes, with charge
${Q}$, and with
masses ${M} {\geq} {Q}$.

The first property is that quantum mechanics and locality/causality
apparently imply an infinite number of internal states of such a black
hole. (See {\it e.g.} \refs{\StTr}.)
To see this, note that although these black holes evaporate, one
can argue that the radiation should shut off as the extremal mass, ${M}
= {Q}$, is approached. (Here the temperature vanishes.)  Now
consider taking such an extremal black hole and
throwing the Earth into it. This will raise it above extremality, but it
subsequently evaporates back to ${M} = {Q}$. A causality argument like
that given for neutral black holes implies that the information carried
by the Earth can't escape. Validity of quantum mechanics implies it
isn't destroyed. Therefore the resultant extremal black hole has
augmented information. Repeating this experiment we can load the black
hole with arbitrarily large information, so it must have an infinite
number of internal states.

The second property is that the Universe is not observed to be unstable
to infinite production of these black holes.

If we can understand how these statements are reconciled for
Reissner-Nordstrom black holes, then they could give us a prototype for
a sensible theory of remnants. On the other hand, failure
to reconcile these would suggest that either quantum mechanics or
locality/causality fails. Pair production of Reissner-Nordstrom black
holes is therefore a litmus test for remnants.

\ifig{\Fig\Schpr}{Above the slice $S$ (dotted) are shown the lorentzian
trajectories of a pair of oppositely charged particles in an electric
field.  Below $S$ is the euclidean continuation of
this solution.  Matching the euclidean and lorentzian solutions smoothly
along $S$ gives a picture of Schwinger production followed
by subsequent evolution of the pair of created
particles.}{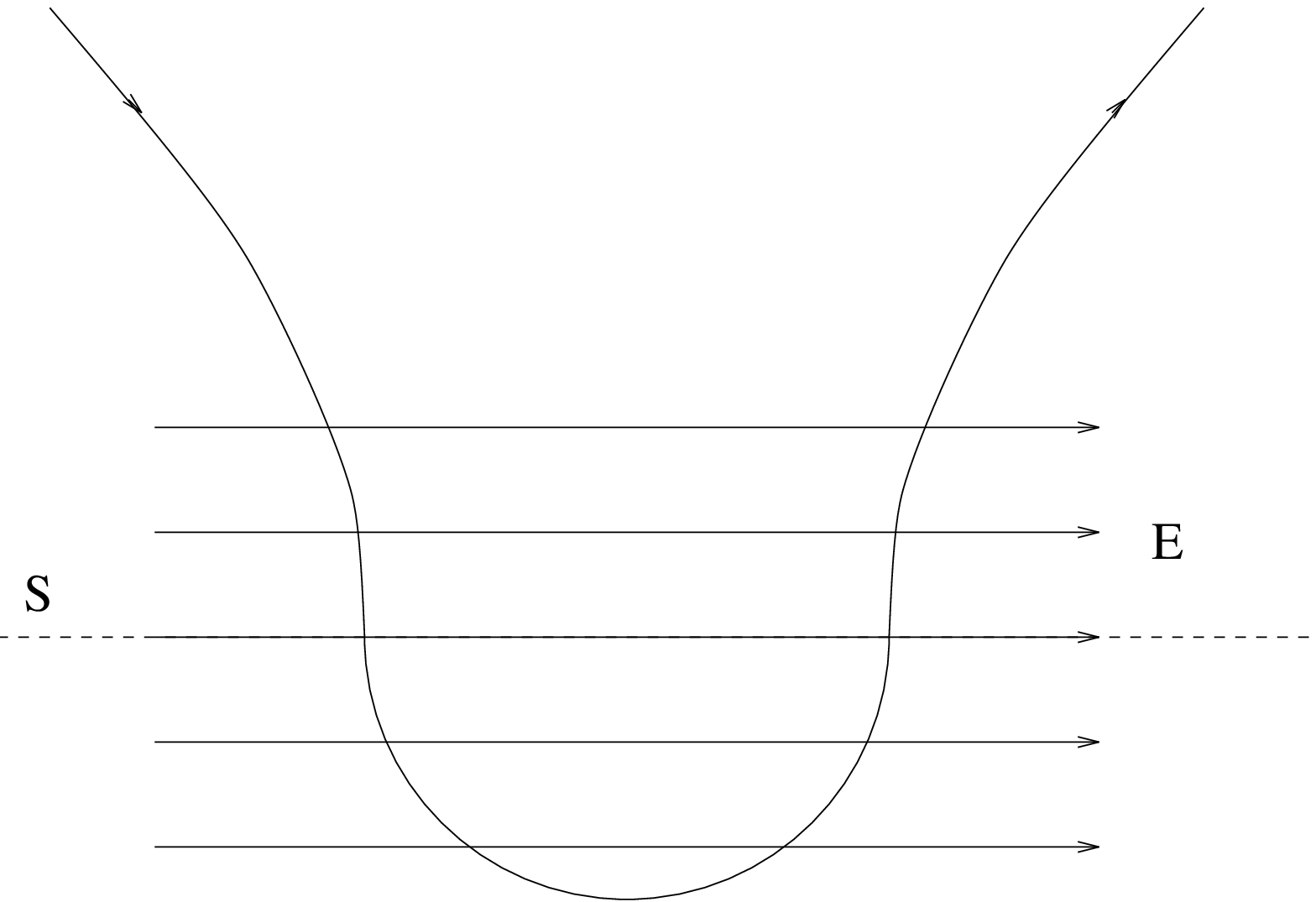}{1.75}

It is easy to investigate such pair production in a uniform background
electromagnetic field. First, recall that Schwinger \refs{\Schw}
calculated the rate of pair production of charged particles of mass
${m}$ and charge ${q}$ in a background electric field and found
\eqn\prrate{
{\Gamma} {\propto} {e^{-\pi m^2/qE}}.}
This can be easily derived by instanton methods. The particles are
produced at rest, and subsequently run off to opposite ends of the
electric field following hyperbolic spacetime trajectories,
as shown in fig.~4.
(By energy conservation, they must be produced with a separation ${\ell}
= {{2m}\over{qE}}.$) The continuation to euclidean time, ${t}{\rightarrow}{i}
{\tau}$, of such trajectories gives a circle, shown in the lower
half of fig.~4. The action of this euclidean trajectory is
\eqn\seven{
{S_{euc}} = {-} {{\pi m^2}\over{qE}},}
which gives the exponent in \prrate.

\ifig{\Fig\spgeom}{The spatial geometry
of a charged black hole.  As the black hole
nears the extremal mass, $M=Q$, the length of the throat region diverges.}
{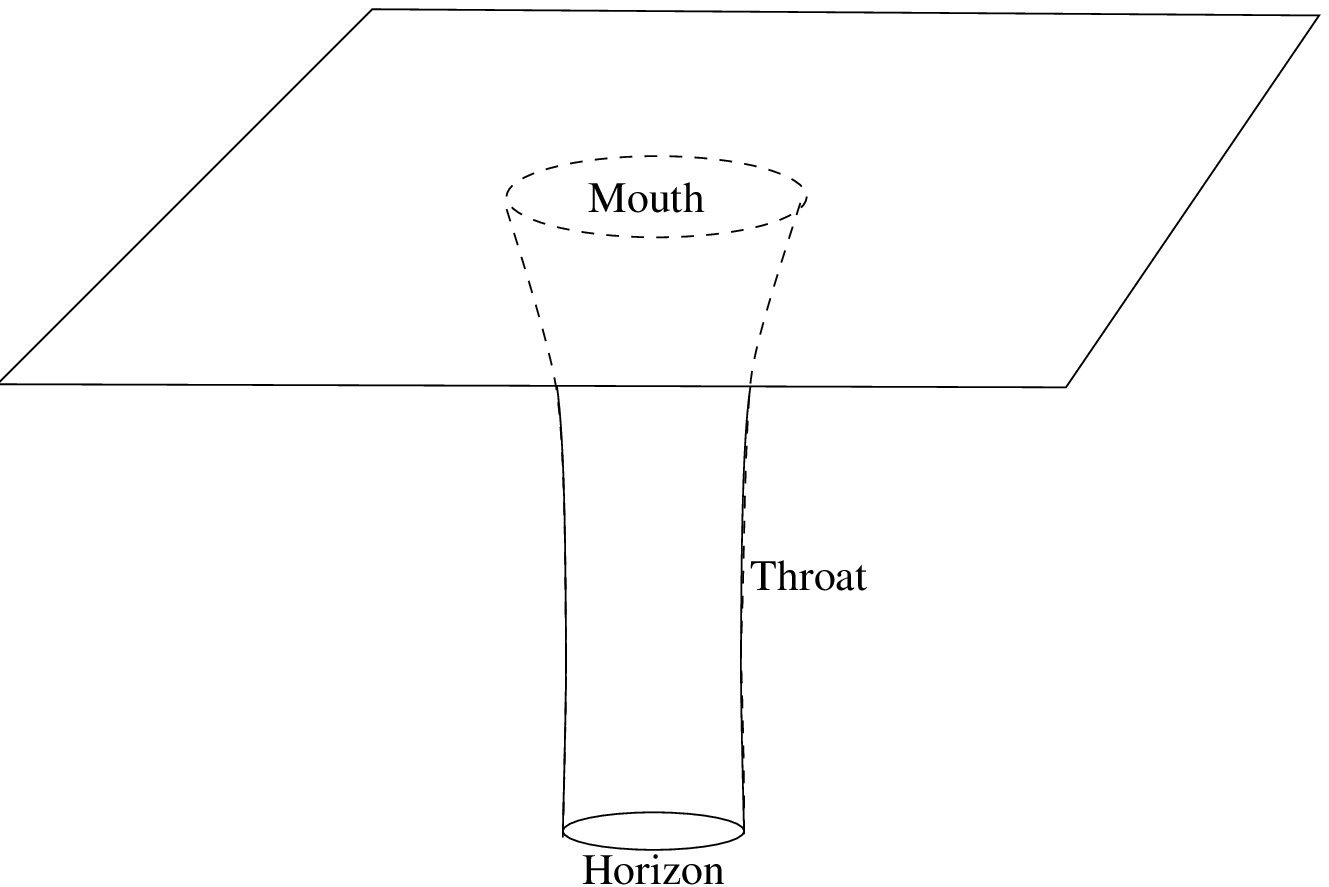}{2.00}

The analogous solution for charged black holes in a background field is
known and is called the Ernst solution \refs{\Ernst}. To avoid problems
of black hole discharge we'll take the charge and field to be magnetic.
The spatial geometry of such a black hole is sketched in fig.~5. There
is a long ``throat'' that connects the outside world to the horizon.
Fig.~6 shows a representation of the Ernst solution, with hyperbolic
motion of the oppositely-charged pair of black holes.

\ifig{\Fig\Lern}{The Ernst solution, which corresponds to a pair of oppositely
charged black holes accelerating in a background field.  The ``fins''
represent the trajectories of the throats of fig.~5, and their
cross-sections are therefore two-spheres.}{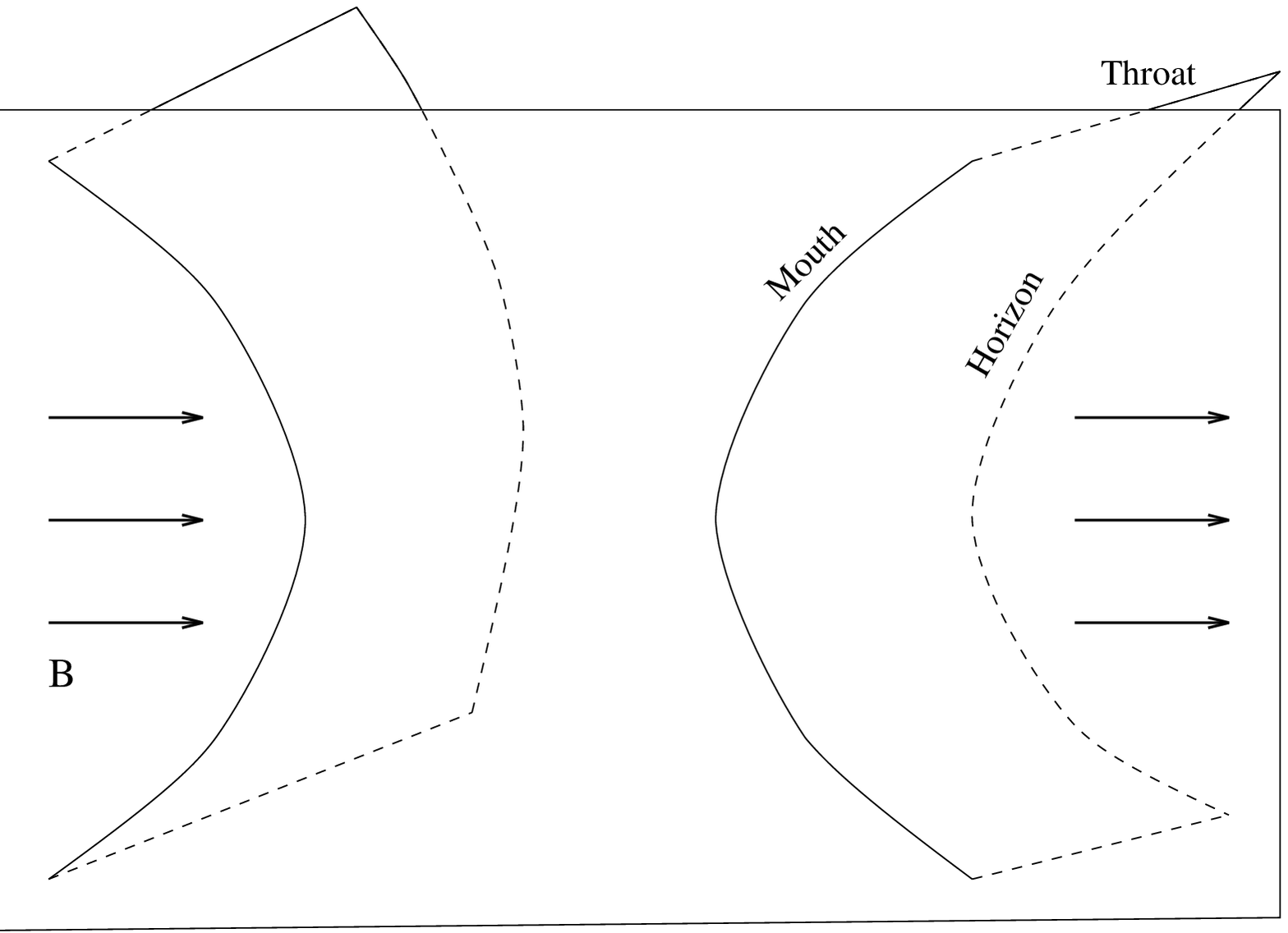}{2.75}

As pointed out by Gibbons \refs{\Gibb} and by Garfinkle and Strominger
\refs{\GaSt}, the euclidean continuation of this solution describes pair
production of black holes. Half of the euclidean
solution is pictured in fig.~7.
Note that the throats appear to connect at the horizon. The production
rate from this configuration is again
\eqn\bhrate{
{\Gamma} {\propto} {e^{-S_{euc}}},}
and a calculation \refs{\GaSt,\GGS} shows that
\eqn\bhact{
{S_{euc}} \simeq {-} {\pi}{Q}/{B} }
in agreement with \prrate.

\ifig{\Fig\EErn}{The euclidean
version of the Ernst solution is the analog of the
lower half of fig.~4.  If one wishes to have a smooth geometry,
the throat regions get identified at the horizon.  This can however be
substantially modified by quantum corrections.}{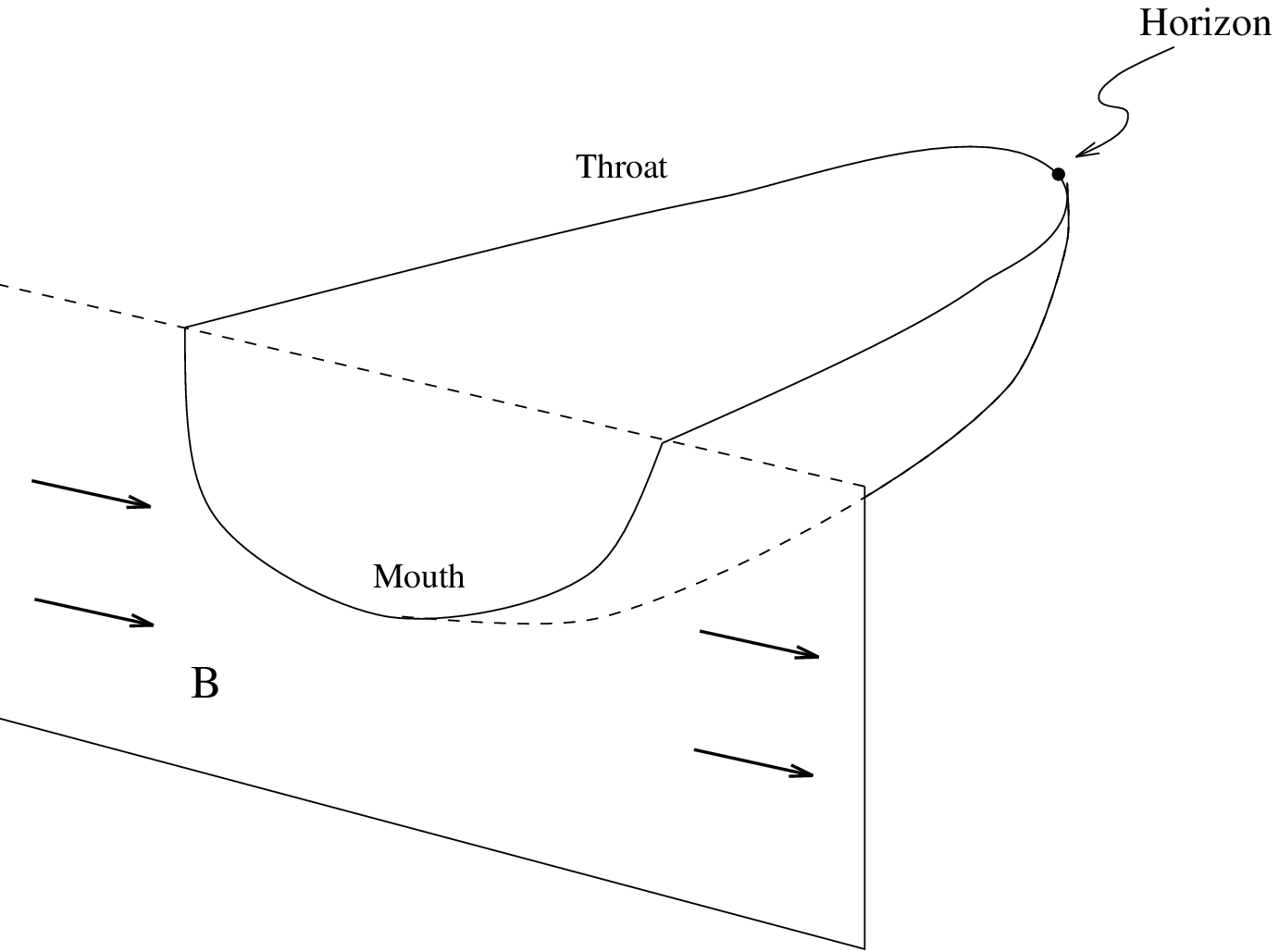}{3.00}

One detail so far omitted is that in both the particle and black hole
case, one should include quantum-mechanical configurations near to the
classical configurations pictured. In the particle case, functional
integration over these configurations gives the known fluctuation
determinant, which contributes
subleading quantum corrections to the rate.
In the black hole case, one is instead integrating over
nearby geometries as well as field configurations, and the result is
hard to compute.

This matter brings us back to the infinite number of states. As we all
know, an astronaut thrown into a black hole appears to get frozen at the
horizon. Therefore, an external observer might describe the infinite
number of black hole states as frozen states at the horizon. When we
integrate over the
neighboring configurations, our integral should include these
configurations in particular.

The states appear to have arbitrarily short wavelengths, and ordinarily
one might expect their contribution to the integral to be suppressed
through \bhrate\ because these wild fluctuations (at least in the frame
of the external observer) would have large euclidean actions. But not so
in gravity, where the negative gravitational energy leads to
cancellations. In fact these configurations have actions of order
\bhact. Integrating over this infinite number of contributions therefore
gives a rate
\eqn\eight{
{\Gamma} = {\infty} {e^{-\pi Q/B}} = {\infty}.}

This reasoning can be checked by examining the limit ${QB}\ll{1}$. In
this limit, the length of the throat region in fig.~7 diverges,
${\ell}\,\,{\sim}\,\,{-}
{Q}{\ell}{n}{Q}{B}$. If one examines the geometry far down this throat,
it appears insensitive to the boundary conditions in the asymptotic
region, aside from knowing about the acceleration due to the field.
Indeed, in this
region, the solution is closely approximated by a free euclidean black
hole. The periodicity corresponds to the fact that it is thermally
excited to a temperature,
\eqn\nine{
{T} {\sim} {B}}
which can be thought of as arising from the acceleration radiation.

The functional integral over the euclidean geometry of a
free black hole
has another
interpretation, as argued by Gibbons and Hawking \refs{\GiHa}: it gives
the partition function for the black hole,
\eqn\ten{
{Z} = {Tr} ({e^{-\beta H}}),}
at the appropriate temperature.  Thus we have what amounts to a ``low-energy
theorem:''  the rate is proportional to the partition function for the
black hole, although we don't know how to explicitly calculate either.
Despite this,
we know that if there are an infinite number of black
hole states with approximately equal energies, \ten\ must be infinite.
This confirms the infinite production rate.

These general arguments, although not completely rigorous, indicate that
the assumption of an infinite number of black hole states indeed implies
infinite pair production. If true this must mean one of our original
assumptions was wrong: either quantum mechanics or locality/causality
fails. This removes the {\it raison d' \^etre} for remnants.

I'll conclude by summarizing my views, although be forewarned that
controversy exists. First, it seems to me fairly clear that there is a
connection between repeatable information loss and energy
non-conservation (although we are investigating this in more detail
\refs{\GiTa}). Furthermore, basic quantum principles would seemingly
force this to occur on a massive scale if it occurs in black hole
evaporation. Thus I see little hope for a resolution of the paradox
here. Second, the  naive expectation of infinite pair production of
remnants seems to be borne out by study of production of charged black
holes. So I don't see much hope here either.

Finally, people have for some time known that comparing observations
inside a black hole to those outside requires comparing reference frames
with ultrahigh relative boosts. This may open a loophole through which
the reasoning behind the paradox fails in a way contrary to our low
energy intuition.  There have been suggestions \refs{\tHoo\SVV-\Sussetal}
that it is locality that fails, and Susskind, Thorlacius, and Uglum have in
particular advocated finding the requisite
non-locality from string theory. Although considerable effort has been
expended to confirm this, and there are suggestive hints, it has not yet
been possible to really test this assertion \refs{\PSTU,\Pol}. The paradox
remains, although it may have just begun to crumble.

\listrefs

\end